\documentclass[a4paper]{ifacconf}

\usepackage{graphicx}      
\usepackage[round]{natbib}        
\usepackage{amsmath,amssymb,bm,mathtools,amsfonts}
\usepackage{cases}
\usepackage{tikz}
\usepackage{xcolor}
\usepackage[inline]{enumitem}

\usepackage{eso-pic}
\AddToShipoutPictureBG*{%
\AtPageUpperLeft{%
\setlength\unitlength{1in}%
\hspace*{\dimexpr0.5\paperwidth\relax}
\makebox(0,-2)[c]{
\begin{tabular}{c c}
T.~J.~{Meijer} \emph{et al.},
Frequency-Domain Data-Driven Predictive Control, \\
To appear in {\em 8th IFAC Conference on Nonlinear Model Predictive Control}, uploaded to ArXiv \today \\
\end{tabular}}}}

\newcommand{\hop}{\mathsf{H} }
\newcommand{\rank}{\operatorname{rank} }

\DeclareMathOperator*{\minimize}{minimize}

\newcounter{assumption}
\newcounter{theorem}
\newcounter{definition}
\newcounter{remark}
\newcounter{lemma}

\newcommand{\new}[1]{{\color{blue} #1}}
\renewcommand{\new}[1]{{#1}}

\begin{document}

\begin{frontmatter}

    
        \title{Frequency-Domain Data-Driven Predictive Control}
        
        \author[TUe]{T.J.~{Meijer}}, 
        \author[TUe]{S.A.N.~{Nouwens}}, 
        \author[TUe]{K.J.A.~{Scheres}},
        \author[ASML]{V.S.~{Dolk}},
        \author[TUe]{W.P.M.H.~{Heemels}}
        
        \address[TUe]{Department of Mechanical Engineering, Eindhoven University of Technology, Eindhoven, The Netherlands}
        \address[ASML]{ASML, De Run 6665, 5504 DT Veldhoven, The Netherlands}

\thanks[footnoteinfo]{This research received funding from the European Research Council (ERC) under the Advanced ERC grant agreement PROACTHIS, no. 101055384.\\ Corresponding author: T.J.~{Meijer}.}

\begin{abstract}                
    In this paper, we propose a data-driven predictive control scheme based on \new{measured} frequency-domain data of the plant. This novel scheme complements the well-known data-driven predictive control (DeePC) approach based on time series data. \new{To develop this new frequency-domain data-driven predictive control (FreePC) scheme, we introduce a novel version of Willems' fundamental lemma based on frequency-domain data.} By exploiting frequency-domain data, we allow recent direct data-driven (predictive) control methodologies to benefit from the available expertise and techniques for non-parametric frequency-domain identification in academia and industry. We prove that, under appropriate conditions, the new FreePC scheme is equivalent to the corresponding DeePC scheme. The strengths of FreePC are demonstrated in a numerical case study.
\end{abstract}


\end{frontmatter}
\everypar{\looseness=-1}
\section{Introduction}
Model predictive control (MPC) is amongst the most successful and widely-adopted control techniques~\citep{Mayne2014}, due to its conceptual simplicity, ability to optimize performance, constraint-handling capabilities, and ease of dealing with multi-input multi-output systems. To achieve this, MPC makes use of a prediction model, which is traditionally obtained through intricate first-principle modeling or identification techniques. An attractive alternative is the so-called data-enabled predictive control (DeePC) scheme introduced in~\citep{Coulson2019} and further extended in recent years. DeePC exploits the celebrated Willems' fundamental lemma (WFL)~\citep{Willems2005,vanWaarde2020,Berberich2020} to characterize the behaviour of the to-be-controlled plant directly in terms of previously-collected time-domain data and has received an enormous amount of attention recently~(see, e.g.,~\citep{Coulson2019,Berberich2021,Verhoek2021b}, or, for a recent survey, see~\citep{Verheijen2023}).

The success of DeePC has inspired many important extensions including, e.g., descriptor systems~\citep{Schmitz2022}, linear parameter-varying systems~\citep{Verhoek2021b}, nonlinear systems~\citep{Berberich2022b,Alsati2023,Lazar2023} and stochastic systems~\citep{Pan2023,Breschi2023}. Similarly, extensions of WFL itself have also been developed and continue to be the subject of ongoing research~(see, e.g.,~\citep{Faulwasser2023} for a recent overview of such extensions). Since WFL was originally formulated using time-domain data of the system, all of the data-driven analysis and control methodologies that it has inspired are also based on time-domain data. This seems like a natural choice for data-driven predictive control because MPC works with time-domain models and optimizes and constrains the time-domain response of the plant. However, since classical control methodologies, such as PID control and loop shaping, are generally designed based on \emph{non-parametric} frequency-domain models/data, their widespread success, particularly in industry, has resulted in an enormous amount of frequency-domain expertise as well as control/identification tools. As it stands, the aforementioned (direct) data-driven approaches are unable to benefit from the available frequency-domain expertise, tools and data. \new{Some noteworthy connections between frequency-domain techniques and predictive control are made, for MPC, in~\citep[][]{Burgos2014,Ozkan2012,Shah2013}, where frequency-domain tuning approaches are proposed, and, for data-driven predictive control, in~\citep{Sathyanarayanan2023}, which presents a scheme based on wavelets that are capable of extracting the frequency content from time-domain data. However, generally speaking, both MPC and data-driven predictive control are purely time-domain control techniques, and direct use of frequency-domain data/models of the plant in predictive control is until now not possible.}

\new{To fill the gap mentioned above, we present a version of WFL based on frequency-domain data (see Theorem~\ref{thm:fd-wfl}), such as frequency-response-function (FRF) measurements. This result is instrumental to fill the gap between recent direct data-driven techniques and frequency-domain data of the plant. In particular, we use it to propose a frequency-domain data-driven predictive control scheme}, which is shown to be equivalent to DeePC in the nominal case. By doing so, we can exploit powerful existing non-parametric frequency-domain identification insights and methods~\citep[see, e.g.,][]{Pintelon2012} in the development of data-driven predictive controllers. \new{Interestingly, an earlier frequency-domain version of WFL is available in~\cite{Ferizbegovic2021}, however, it does not exploit the conjugate symmetry due to the underlying system being real and, as a result, it can produce complex-valued trajectories despite the plant being real. We do exploit conjugate symmetry in our version of WFL and show its use in FreePC. \new{Another relevant recent work connects frequency domain and WFL by using WFL to compute the frequency response function based on time-domain data~\citep{Markovsky2024}, which essentially complements our objective of using frequency-domain data to characterize (in time-domain) the dynamics of the system.} To conclude our paper, we demonstrate the potential of our results in a numerical case study.}

The content of the paper is organized as follows. We provide some relevant preliminaries in Section~\ref{sec:prelim} and we introduce the problem statement in Section~\ref{sec:problem-statement}. Section~\ref{sec:freedom-pc} presents our main contributions consisting of a version of WFL based on frequency-domain data and a frequency-domain data-driven predictive control scheme. Finally, a numerical case study to demonstrate our results is included in Section~\ref{sec:case-study} followed by some conclusions in Section~\ref{sec:conclusions}.

\section{Preliminaries}\label{sec:prelim}  
{\bf Notation.} We denote the set of real, integer and non-negative integers by, respectively, $\mathbb{R}$, $\mathbb{Z}$, $\mathbb{N}$. Let $\mathbb{N}_{[n,m]}=\{n,n+1,\hdots,m\}$ and $\mathbb{N}_{\geqslant n}=\{n,n+1,\hdots\}$ with $n,m\in\mathbb{N}$, $\mathbb{W}=[-\pi,\pi)\subset\mathbb{R}$ and $\mathbb{W}_+=[0,\pi)\subset\mathbb{W}$. Let $v\in\mathbb{C}^{n}$, then $v^\top$, $v^\hop$ and $v^*$ denote, respectively, its transpose, its complex-conjugate transpose, and its complex conjugate while $\|v\|_1\coloneq \sum_{i\in\mathbb{N}_{[1,n]}}|v_i|$ denotes its $1$-norm. Moreover, $\operatorname{Re}v$ and $\operatorname{Im}v$ denote the real and imaginary part of $v$ and we denote $j^2 =-1$. The notation $(u,v)$ stands for $\begin{bmatrix}\begin{smallmatrix} u^\top & v^\top\end{smallmatrix}\end{bmatrix}^\top$. We denote sequences using boldface, i.e., $\bm{x}=\{x_k\}_{k\in\mathbb{N}_{[0,N-1]}}$, $N\in\mathbb{N}$, and the notation $x_{[n,m]}=(x_n,x_{n+1},\hdots,x_m)$, $n,m\in\mathbb{N}_{[n,m]}$, is used to denote a vector containing vertically-stacked elements of a part of the sequence $\bm{x}$. Moreover, $\mathcal{R}^{n}_{N}\coloneqq \{\{x_k\}_{k\in\mathbb{N}_{[0,N-1]}}\mid x_k\in\mathbb{R}^{n},k\in\mathbb{N}_{[0,N-1]}\}$, $\mathcal{C}^{n}_{N}\coloneqq \{\{x_k\}_{k\in\mathbb{N}_{[0,N-1]}}\mid x_k\in\mathbb{C}^{n},k\in\mathbb{N}_{[0,N-1]}\}$ and $\mathcal{W}^+_N\coloneqq \{\{\omega_k\}_{k\in\mathbb{N}_{[0,N-1]}}\mid\omega_k\in\mathbb{W}_+,k\in\mathbb{N}_{[0,N-1]}\}$ are the sets of length-$N$ sequences taking values in $\mathbb{R}^{n}$, $\mathbb{C}^{n}$ and $\mathbb{W}_+$, respectively. For any $\bm{u}\in\mathcal{C}^{n}_N$ and $\bm{v}\in\mathcal{C}^{m}_{N}$, we denote $\{\bm{u},\bm{v}\}=\{(u_k,v_k)\}_{k\in\mathbb{N}_{[0,N-1]}}\in\mathcal{C}^{n+m}_N$. Let $H_L(\bm{x})$ for $\bm{x}\in\mathcal{R}^{n}_N$ denote the depth-$L$, with $L\leqslant N$, Hankel matrix induced by $\bm{x}$, i.e.,
\begin{equation*}
    H_L(\bm{x})=\begin{bmatrix} 
        x_{[0,L-1]} & x_{[1,L]} & \hdots & x_{[N-L,N-1]}
    \end{bmatrix}.
\end{equation*}
Finally, $\otimes$ denotes the Kronecker product.

Consider the LTI system $\Sigma$ given by
\begin{subnumcases}{\label{eq:td-system}\Sigma~\colon~}
    x_{k+1} &= $Ax_k + Bu_k,$\label{eq:td-system-state}\\
    y_k &= $Cx_k + Du_k,$\label{eq:td-system-output}
\end{subnumcases}
$x_k\in\mathbb{R}^{n_x}$, $u_k\in\mathbb{R}^{n_u}$ and $y_k,n_k\in\mathbb{R}^{n_y}$ denote, respectively, the state, input and output of $\Sigma$ at time $k\in\mathbb{Z}$. The transfer function $G(z)$ of $\Sigma$ is given by
\begin{equation}
    G(z) = C(zI-A)^{-1}B+D,\quad z\in\mathbb{C}.
\end{equation}
We assume that the pair $(A,B)$ is controllable. In the sequel, we assume that we are given data generated by $\Sigma$, while $\Sigma$ itself is unknown. In DeePC, this data consists of an input-output trajectory with the input being persistently exciting (PE)~\cite[see, e.g.,][]{Berberich2020,vanWaarde2020,Willems2005}.
\setcounter{thm}{\thedefinition}\stepcounter{definition}
\begin{defn}\label{dfn:trajectory}
    A pair of sequences $\{\bm{u},\bm{y}\}$ with $\bm{u}\in\mathcal{R}^{n_u}_{N}$ and $\bm{y}\in\mathcal{R}^{n_y}_{N}$ is called an input-output trajectory of $\Sigma$ in~\eqref{eq:td-system}, if there exists a state sequence $\bm{x}\in\mathcal{R}^{n_x}_{N}$ such that $\{\bm{u},\bm{x},\bm{y}\}$ satisfies~\eqref{eq:td-system-state} for $k\in\mathbb{N}_{[0,N-2]}$ and~\eqref{eq:td-system-output} for $k\in\mathbb{N}_{[0,N-1]}$.
\end{defn}
\setcounter{thm}{\thedefinition}\stepcounter{definition}
\begin{defn}\label{dfn:CPE}
    The sequence $\bm{v}\in\mathcal{R}^{n_v}_{N}$ is said to be persistently exciting (PE) of order $L\in\mathbb{N}_{\geqslant 1}$, with $L\leqslant N$, if
    \begin{equation*}
        \rank H_L(\bm{v})=n_vL.
    \end{equation*}
\end{defn}
Next, we recall WFL.
\setcounter{thm}{\thelemma}\stepcounter{lemma}
\begin{lem}\label{lem:WFL}
    Let $\{\bm{u}^{d},\bm{y}^{d}\}$ with $\bm{u}^{d}\in\mathcal{R}^{n_u}_{N}$ and $\bm{y}^{d}\in\mathcal{R}^{n_y}_{N}$ be an input-output trajectory of $\Sigma$. Suppose that $\bm{u}^{d}$ is PE of order $L+n_x$ with $L\in\mathbb{N}_{\geqslant 1}$. Then, $\{\bm{u},\bm{y}\}$ with $\bm{u}\in\mathcal{R}^{n_u}_{L}$ and $\bm{y}\in\mathcal{R}^{n_y}_{L}$ is an input-output trajectory of $\Sigma$, if and only if there exists $g\in\mathbb{R}^{N-L+1}$ such that\footnote{We use the superscript $d$ to denote data that was collected off-line.}
    \begin{equation*}
        \begin{bmatrix}
            u_{[0,L-1]}\\
            y_{[0,L-1]}
        \end{bmatrix} = \begin{bmatrix}
            H_L(\bm{u}^{d})\\
            H_L(\bm{y}^{d})
        \end{bmatrix}g.
    \end{equation*}
\end{lem}
Lemma~\ref{lem:WFL}, which was introduced in~\citep{Willems2005} and has been the foundation of many recent direct data-driven (predictive) control approaches, states that all possible length-$L$ solutions to~\eqref{eq:td-system} can be characterized using a single data sequence of $N$ input-output data points with the input being persistently exciting. This result is used in DeePC to obtain a prediction model directly in terms of (time-domain) data. Since the (internal) state of the system is unknown, an initial input-output trajectory is prepended to the predicted input-output trajectories to enforce consistency (of the predictions) with the unknown internal state of the system. Note that, since we do not assume observability of $\Sigma$, this internal state may not be uniquely determined based on an initial input-output sequence, however, by using an initial sequence of length $\bar{T}\in\mathbb{N}_{\geqslant 1}$ with $\bar{T}\geqslant n_x$, the observable part of the state is unique and, as a result, the future output is, for any given input sequence, also unique. Let $T\in\mathbb{N}_{\geqslant 1}$ be the prediction horizon and let $\{\bm{u}^d,\bm{y}^d\}$ with $\bm{u}^d\in\mathcal{R}^{n_u}_{N}$ and $\bm{y}^d\in\mathcal{R}^{n_y}_{N}$ be an input-output sequence satisfying the following:
\setcounter{thm}{\theassumption}\stepcounter{assumption}
\begin{assum}\label{asm:PE-td}
    The sequence $\bm{u}^d$ is PE of order $\bar{T}+T$ with $\bar{T}\geqslant n_x$.    
\end{assum}
Then, given two vectors of stacked past input/output data $\bar{u}_k=u_{[k-\bar{T},k-1]}$ and $\bar{y}_k=y_{[k-\bar{T},k-1]}$, DeePC solves the following finite-horizon optimal control problem at every time $k\in\mathbb{N}$:
\begin{equation}\setlength{\arraycolsep}{0.5em}
	\begin{array}{@{}cl@{}} \minimize_{\bm{u}_k,\bm{y}_k,g_k,\sigma_k} & \lambda_\sigma\|\sigma_k\|_1+\lambda_g\|g_k\|_1+\sum_{\mathclap{i\in\mathbb{N}_{[0,T-1]}}} \ell(y_{i,k},u_{i,k})\\
	\text{subject to} & \left[\begin{array}{@{}c@{}}
        \bar{u}_k\\
        u_{[0,T-1],k}\\\hline
        \bar{y}_k+\sigma_k\\
        y_{[0,T-1],k}
    \end{array}\right] = \left[\begin{array}{@{}c@{}}
        H_{\bar{T}+T}(\bm{u}^d)\\\hline
        H_{\bar{T}+T}(\bm{y}^d)
    \end{array}\right]g_k,\\
    & u_{i,k}\in\mathbb{U},~y_{i,k}\in\mathbb{Y},~\text{ for all }i\in\mathbb{N}_{[0,T-1]},
    \end{array}\label{eq:DeePC}
\end{equation}
where $\mathbb{U}$ and $\mathbb{Y}$ denote the set of admissible inputs and outputs, respectively, while $\ell\colon\mathbb{R}^{n_y}\times\mathbb{R}^{n_u}\rightarrow\mathbb{R}_{\geqslant 0}$ denotes the stage cost. The decision variables in~\eqref{eq:DeePC} are $g_k\in\mathbb{R}^{N-T-\bar{T}+1}$, $\sigma_k\in\mathbb{R}^{\bar{T}n_y}$, $y_{i,k}$ and $u_{i,k}$, $i\in\mathbb{N}_{[0,T-1]}$, where denote, respectively, the prediction of $y_{k+i}$ and $u_{k+i}$, respectively, computed at time $k$. Here, $\sigma_k$ is an auxiliary slack variable and $\lambda_\sigma,\lambda_g\in\mathbb{R}_{>0}$ are regularization parameters~\citep[][]{Coulson2019}. These regularization parameters and slack variables are needed to deal with noise in the offline data as well as the past output trajectory $\bar{y}_k$\new{, as demonstrated in Section~\ref{sec:case-study}}. In the noiseless/nominal case, however, we can use $\lambda_g=0$ and $\sigma_k=0$. Moreover, $y_{[0,T-1],k}=(y_{0,k},y_{1,k},\hdots,y_{T-1,k})$ denotes the vector obtained by stacking the predicted output at time $k$ and, similarly, $u_{[0,T-1],k} = (u_{0,k},y_{1,k},\hdots,y_{T-1,k})$. Let $\bm{u}^\star_k=(u^\star_{0,k},u^\star_{1,k},\hdots,u^\star_{T-1,k})$ denote the optimal control action computed at time $k$ by solving~\eqref{eq:DeePC}. DeePC implements the first element of $\bm{u}^\star_k$, i.e., $u_k=u^\star_{0,k}$, and solves~\eqref{eq:DeePC} at time $k+1$ using the vectors of stacked past input/output data $\bar{u}_{k+1}=u_{[k-\bar{T}+1,k]}$ and $\bar{y}_{k+1}=y_{[k-\bar{T}+1,k]}$. 

\section{Problem statement}\label{sec:problem-statement}
As mentioned before, DeePC and all of its extensions and/or generalizations are formulated using time-domain data, i.e., the Hankel matrices in~\eqref{eq:DeePC} are constructed directly in terms of previously-collected time-domain measurements of $\Sigma$. However, any time-domain sequence $\bm{v}=\{v_k\}_{k\in\mathbb{Z}}$, with $v_k\in\mathbb{R}^{n_v}$ for all $k\in\mathbb{Z}$, can be fully characterized in frequency domain by its spectrum $V(\omega)$, $\omega\in\mathbb{W}$, given by the discrete-time Fourier transform (DTFT)
\begin{equation}\label{eq:DTFT}
    V(\omega) = \sum_{k=-\infty}^\infty v_ke^{-j\omega k},\quad \omega\in\mathbb{W}.
\end{equation}
\new{Our objective, in this paper, is to formulate a frequency-domain data-driven predictive control scheme in which the offline data consists of samples of the input spectrum $U(\omega)$ and the corresponding output spectrum $Y(\omega)$ instead of the time-domain data $\bm{u}^d$ and $\bm{y}^d$ used in DeePC. To facilitate this, we aim to develop a frequency-domain counterpart to WFL. Finally, we investigate the equivalence between DeePC and FreePC in the nominal/noiseless case (i.e., $\lambda_g=0$ and $\sigma_k=0$).}


\section{Main results}\label{sec:freedom-pc}
\new{In this section, we subsequently introduce a version of WFL based on frequency-domain data and use it to formulate our FreePC scheme.}

\new{\subsection{Frequency-domain Willems' fundamental lemma}}
The time-domain signal $\bm{v}=\{v_k\}_{k\in\mathbb{Z}}$, which can be obtained from its spectrum $V(\omega)$ using the inverse DTFT
\begin{equation*}
    v_k = \frac{1}{2\pi}\int_\mathbb{W} V(\omega)e^{j\omega k},\quad k\in\mathbb{Z},
\end{equation*}
is real-valued if and only if $V(\omega)$ is symmetric, i.e., $V(\omega)=V^*(-\omega)$ for all $\omega\in\mathbb{W}$. By taking the DTFT of~\eqref{eq:td-system}, we can alternatively define solutions to $\Sigma$ in the frequency domain~\citep{Hespanha2018}.
\setcounter{thm}{\thedefinition}\stepcounter{definition}
\begin{defn}\label{dfn:input-output-spectrum}
    A pair of spectra $\{U(\omega),Y(\omega)\}$ with $U(\omega)=U^*(-\omega)\in\mathbb{C}^{n_u}$ and $Y(\omega)=Y^*(-\omega)\in\mathbb{C}^{n_y}$, $\omega\in\mathbb{W}$, is said to be an input-output spectrum of $\Sigma$, if
    \begin{equation}
        Y(\omega)=G(e^{j\omega})U(\omega),\quad \text{for all }\omega\in\mathbb{W}.\label{eq:input-output-spectrum}
    \end{equation}
\end{defn}
If~\eqref{eq:input-output-spectrum} holds, then $X(\omega)=X^*(-\omega)=(e^{j\omega}I-A)^{-1}BU(\omega)$, $\omega\in\mathbb{W}$, is a state spectrum of $\Sigma$ in the sense that its inverse DTFT satisfies~\eqref{eq:td-system}. 

Since it is not feasible to measure an input-output spectrum at all frequencies in $\mathbb{W}$, we will use $M\in\mathbb{N}_{\geqslant 1}$ samples of $U^{d}(\omega)$ and $Y^{d}(\omega)$ at the frequencies $\bm{\omega}\in\mathcal{W}^+_M$. These samples are collected in the complex-valued sequences $\bm{U}^{d}\in\mathcal{C}^{n_u}_M$ and $\bm{Y}^{d}\in\mathcal{C}^{n_y}_M$, respectively. Since we can exploit the symmetry in the spectra, we only consider frequencies between $[0,\pi)$, i.e., $\bm{\omega}\in\mathcal{W}^+_M$. \new{Note that we \begin{enumerate*}[label=(\alph*)] \item do not require any processing of the data (other than taking the DTFT, which is invertible), and \item do not impose any structure on $\Sigma$ (aside from linearity assumed in~\eqref{eq:td-system}). \end{enumerate*}} As we will illustrate in Section~\ref{sec:case-study}, for periodic time-domain sequences (e.g., when periodic excitation is used and the free response has damped out), the infinite sum in the DTFT~\eqref{eq:DTFT} can be computed based on finite-length data containing an integer number of periods of the sequence.
\setcounter{thm}{\theremark}\stepcounter{remark}
\begin{rem}\label{rem:frf-meas}
    These sampled input-output spectra also accommodate the important case, where we are given $M$ measurements of the transfer function in a specific input direction, i.e., we are given $G(e^{j\omega_m})r_m$, for $m\in\mathcal{M}$, with $r_m$ being the $m$-th input direction. This can be incorporated by setting $Y^d_m = G(e^{j\omega_m})r_m$ and $U^d_m=r_m$, $m\in\mathcal{M}$. \new{Note that we do not need to identify the transfer function to do this and we do not impose any specific model structure other than the linearity in~\eqref{eq:td-system}.} 
\end{rem}
Naturally, to fully characterize $\Sigma$ based on sampled input-output spectra, we require the data to be sufficiently ``rich''. To formalize this, we repeat here the frequency-domain persistence of excitation notion from~\citep{Meijer2024-ecc-arxiv}. Let $F_L(\bm{V},\bm{\omega})\in\mathbb{C}^{n_vL\times M}$, $L\in\mathbb{N}$, for any $\bm{V}\in\mathcal{C}^{n_v}_M$ and $\bm{\omega}\in\mathcal{W}^+_M$, $M\in\mathbb{N}$, be given by \begin{equation*}
    F_L(\bm{V},\bm{\omega}) \coloneqq \begin{bmatrix}
        W_L(\omega_0)\otimes V_0 & \hdots W_L(\omega_{M-1})\otimes V_{M-1}
    \end{bmatrix}, 
\end{equation*}
where $W_L(\omega)\coloneqq \begin{bmatrix}1 & e^{j\omega} & \hdots & e^{j\omega(L-1)}\end{bmatrix}^\top$, $\omega\in\mathbb{W}$. 
\setcounter{thm}{\thedefinition}\stepcounter{definition}
\begin{defn}\label{dfn:fd-PE}
    Consider the sequence $\bm{V}\in\mathcal{C}^{n_v}_M$, $M\in\mathbb{N}_{\geqslant 1}$, containing samples of the spectrum $V(\omega)$, $\omega\in\mathbb{W}$, at the frequencies $\bm{\omega}\in\mathcal{W}^+_M$, i.e., $V_m=V(\omega_m)$ for $m\in\mathcal{M}$. Then, $\bm{V}$ is said to be PE of order $L\in\mathbb{N}_{\geqslant 1}$, if
    \begin{equation*}
        \rank \begin{bmatrix} F_L(\bm{V},\bm{\omega}) & F_L^*(\bm{V},\bm{\omega})\end{bmatrix}=n_vL.
    \end{equation*}
\end{defn}
In~\citep[][]{Colin2020}, a constructive frequency-domain method was proposed, for MIMO systems, to design signals of a desired PE order. \new{We are now ready to introduce our frequency-domain version of WFL below.}
\setcounter{thm}{\thetheorem}\stepcounter{theorem}
\begin{thm}\label{thm:fd-wfl}
    Consider the system $\Sigma$ in~\eqref{eq:td-system} and the sequences $\bm{U}^{d}\in\mathcal{C}^{n_u}_M$ and $\bm{Y}^{d}\in\mathcal{C}^{n_y}_M$, $M\in\mathbb{N}_{\geqslant 1}$, containing samples of the input-output spectrum $\{U^d(\omega),Y^d(\omega)\}$ at the frequencies $\bm{\omega}\in\mathcal{W}^+_M$, i.e., $U^d_m=U^d(\omega_m)$ and $Y^d_m=Y^d(\omega_m)$ for $m\in\mathcal{M}$. Suppose that $\bm{U}^{d}$ is PE of order $L+n_x$. Then, $\{\bm{u},\bm{y}\}$, with $\bm{u}\in\mathcal{R}^{n_u}_L$ and $\bm{y}\in\mathcal{R}^{n_y}_L$, is an input-output trajectory of $\Sigma$ if and only if there exists $g\in\mathbb{R}^{2M}$ such that
    \begin{equation}
        \begin{bmatrix}
            u_{[0,L-1]}\\\hline
            y_{[0,L-1]}
        \end{bmatrix} = \left[\begin{array}{@{}cc@{}}
            \operatorname{Re} F_L(\bm{U}^d,\bm{\omega}) & \operatorname{Im} F_L(\bm{U}^d,\bm{\omega})\\\hline
            \operatorname{Re} F_L(\bm{Y}^d,\bm{\omega}) & \operatorname{Im} F_L(\bm{Y}^d,\bm{\omega})
        \end{array}\right]g.
    \end{equation}\label{eq:fd-WFL}
\end{thm}
\new{For the proof of Theorem~\ref{thm:fd-wfl}, we refer to~\cite{Meijer2024-ecc-arxiv}, where a more general version of Theorem~\ref{thm:fd-wfl}, in which the input-output spectrum may be sampled using multiple input directions at the same frequency, is proved. Both Definition~\ref{dfn:fd-PE} and Theorem~\ref{thm:fd-wfl} exploit the symmetry of $V(\omega)$. In Definition~\ref{dfn:fd-PE} this is done by including $F_L^*(\bm{V},\bm{\omega})$, while in Theorem~\ref{thm:fd-wfl} the conjugate symmetry has been further exploited to arrive at the real and imaginary parts in~\eqref{eq:fd-WFL}. Exploiting conjugate symmetry results in up to $2$ orders of PE per excited frequency and it ensures that the left-hand side of~\eqref{eq:fd-WFL} is always real-valued. Interestingly, the latter is not true for the frequency-domain WFL presented in~\cite{Ferizbegovic2021}, which can result in complex-valued solutions despite $\Sigma$ being real.} Theorem~\ref{thm:fd-wfl} allows only one input direction per frequency to be used. As a result, if we are given measurements of $G(e^{j\omega_m})$ as a whole, we are unable to exploit all of the available data. To overcome this,~\citep{Meijer2024-ecc-arxiv} extends Theorem~\ref{thm:fd-wfl} to allow multiple input directions per frequency. 

\new{\subsection{FreePC}}
We are now ready to present our so-called FreePC scheme, which is the frequency-domain data-driven counterpart to DeePC~\eqref{eq:DeePC} and forms the main contribution of this paper. Let $\bar{T}\in\mathbb{N}_{\geqslant n_x}$ be the length of the past input-output data, let $T\in\mathbb{N}_{\geqslant 1}$ be the prediction horizon and let $\{\bm{U}^d,\bm{Y}^d\}$, with $\bm{U}^d\in\mathcal{C}^{n_u}_M$ and $\bm{Y}^{d}\in\mathcal{C}^{n_y}_{M}$, be an input-output sequence satisfying the following:
\setcounter{thm}{\theassumption}\stepcounter{assumption}
\begin{assum}\label{asm:PE-fd}
    $\bm{U}^d$ is PE of order $\bar{T}+T$ with $\bar{T}\geqslant n_x$.
\end{assum}
Then, given two vectors of stacked past input/output data $\bar{u}_k$ and $\bar{y}_k$, FreePC solves, at every time $k\in\mathbb{N}$, the finite-horizon optimal control problem
\begin{multline}\setlength{\arraycolsep}{0.5em}
	\begin{array}{@{}cl@{}} \minimize_{\bm{u}_k,\bm{y}_k,g_k,\sigma_k} & \lambda_\sigma \|\sigma_k\|_1+\lambda_g\|g_k\|_1+\sum_{\mathclap{i\in\mathbb{N}_{[0,T-1]}}} \ell(y_{i,k},u_{i,k})\\
	\text{subject to} & \left[\begin{array}{@{}c@{}}
        \bar{u}_k\\
        u_{[0,T-1],k}\\\hline
        \bar{y}_k+\sigma_k\\
        y_{[0,T-1],k}
    \end{array}\right] = \\
    &\quad \left[\begin{array}{@{}cc@{}}
        \operatorname{Re} F_{\bar{T}+T}(\bm{U}^d,\bm{\omega}) & \operatorname{Im} F_{\bar{T}+T}(\bm{U}^d,\bm{\omega})\\\hline
        \operatorname{Re} F_{\bar{T}+T}(\bm{Y}^d,\bm{\omega}) & \operatorname{Im} F_{\bar{T}+T}(\bm{Y}^d,\bm{\omega})
    \end{array}\right]g_k,\\
    \end{array}\\
    u_{i,k}\in\mathbb{U},~y_{i,k}\in\mathbb{Y},~\text{ for all }i\in\mathbb{N}_{[0,T-1]}.
    \label{eq:FreePC}
\end{multline}
After solving~\eqref{eq:FreePC}, we implement the first element of the optimal control action $\bm{u}^\star_k$, i.e., $u_k=u^\star_{0,k}$. At time $k+1$, this process is repeated based on $\bar{u}_{k+1}$ and $\bar{y}_{k+1}$. In~\eqref{eq:FreePC}, similar to DeePC~\eqref{eq:DeePC}, the stage cost is denoted by $\ell$, the sets of admissible inputs and outputs, respectively, by $\mathbb{U}$ and $\mathbb{Y}$, $\sigma_k\in\mathbb{R}^{\bar{T}n_y}$ is an auxiliary slack variable and $\lambda_\sigma,\lambda_g\in\mathbb{R}_{>0}$ are regularization parameters.

Observe that the key difference between FreePC and DeePC is in the models describing $(\bar{u}_k,u_{[0,T-1],k},\bar{y}_k,\allowbreak y_{[0,T-1],k})$. In DeePC, this model is formulated using the time-domain data $\{\bm{u}^d,\bm{y}^d\}$, whereas FreePC uses the frequency-domain data $\{\bm{U}^d,\bm{Y}^d\}$. It is worth mentioning that, although the matrices $F_{\bar{T}+T}(\bm{U}^d,\bm{\omega})$ and $F_{\bar{T}+T}^d(\bm{Y}^{d},\bm{\omega})$ are complex-valued, the model used in FreePC is real-valued (since we take the real and imaginary parts). As such, $g_k\in\mathbb{R}^{2M}$ is real-valued as well. As mentioned in Remark~\ref{rem:frf-meas},~\eqref{eq:FreePC} allows only one input direction in the data $\bm{U}^d$ to be excited at each frequency, which can be extended using the approach in~\citep{Meijer2024-ecc-arxiv}. 

\new{Next, we show that in the nominal case (where $\lambda_g=0$ and $\sigma_k=0$, which is possible when the offline data and $\bar{y}_k$ are noise-free) FreePC is equivalent to DeePC~\eqref{eq:DeePC}.}
\setcounter{thm}{\thetheorem}\stepcounter{theorem}
\begin{thm}\label{thm:equiv}
    Consider the system $\Sigma$ in~\eqref{eq:td-system} and let $\lambda_g=0$ and $\sigma_k=0$ in~\eqref{eq:DeePC} and~\eqref{eq:FreePC}. Then, FreePC based on~\eqref{eq:FreePC} is equivalent to DeePC based on~\eqref{eq:DeePC} in the sense that, at any $k\in\mathbb{N}$, given the same vectors of stacked past input/output data $\bar{u}_k$ and $\bar{y}_k$, the open-loop optimal control problems solved in~\eqref{eq:FreePC} and~\eqref{eq:DeePC} admit the same (set of) optimal control and corresponding output sequence(s).
\end{thm}
\new{The proof of Theorem~\ref{thm:equiv} follows from Lemma~\ref{lem:WFL} and Theorem~\ref{thm:fd-wfl} and is omitted for space reasons.} It is immediate from Theorem~\ref{thm:equiv} that, if~\eqref{eq:DeePC} admits a unique optimal control action $u^\star_{0,k}$, then $u^\star_{0,k}$ is also the unique optimal control action obtained by solving~\eqref{eq:FreePC} and vice versa.
\setcounter{thm}{\theremark}\stepcounter{remark}
\begin{rem}\label{rem:complexity}
    In DeePC~\eqref{eq:DeePC}, the number of decision variables depends on the length of the time-domain data, since $g_k\in\mathbb{R}^{N-T-\bar{T}+1}$ in~\eqref{eq:DeePC}, where $N$ denotes the length of the offline (time-domain) data. Interestingly, in FreePC~\eqref{eq:FreePC}, the number of decision variables only depends on the number of frequencies in $\bm{\omega}$\new{, i.e., the number of frequencies at which we sample the input-output spectrum}, and not on the length of the time-domain data used to determine the frequency-domain sequences $\bm{U}^d$ and $\bm{Y}^d$. As we will illustrate in the next section, we can exploit this fact to perform longer experiments to collect more offline data and, thereby, reduce the effect of noise, without increasing the computational complexity of FreePC.
\end{rem}

\section{Numerical case study}\label{sec:case-study}
\begin{figure}[!t]
    \centering
    \includegraphics[height=.3\linewidth]{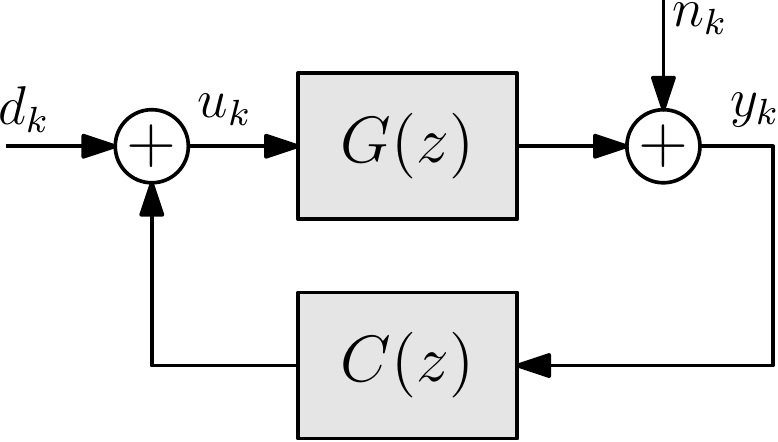}
    \caption{Closed-loop measurement setup.}
    \label{fig:closed-loop-meas}
\end{figure}

We consider the unstable single-input single-output (SISO) system $\Sigma$ of the form~\eqref{eq:td-system} with transfer function
\begin{equation}
    G(z) = \frac{0.1164z+0.1071}{z^2 - 1.891z + 0.7788}.\label{eq:example-G}
\end{equation}
\new{In order to investigate the effect of measurement noise on the proposed FreePC scheme, we replace~\eqref{eq:td-system-output} with 
\begin{equation}
    y_k = Cx_k + Du_k + n_k.
\end{equation}}%
Since the system~\eqref{eq:example-G} is unstable, we collect our offline data in a closed-loop measurement setup~\citep[see, e.g.,][Chapter 10]{Soderstrom1989}, as depicted in Fig.~\ref{fig:closed-loop-meas}, with the stabilizing controller
\begin{equation*}
    C(z) = \frac{6z-5.135}{z-0.1353}.
\end{equation*}
As shown in Fig.~\ref{fig:closed-loop-meas}, the input $u_k\in\mathbb{R}$ consists of the output of the controller on top of which we inject a signal $\{d_k\}_{k\in\mathbb{N}}$ with $d_k\in\mathbb{R}$ for all $k\in\mathbb{N}$. \new{The ease of performing data collection in such a closed-loop setting is, particularly, for unstable systems, another important benefit of considering frequency-domain data.}

\begin{figure}[!t]
    \centering
    \includegraphics[width=\linewidth]{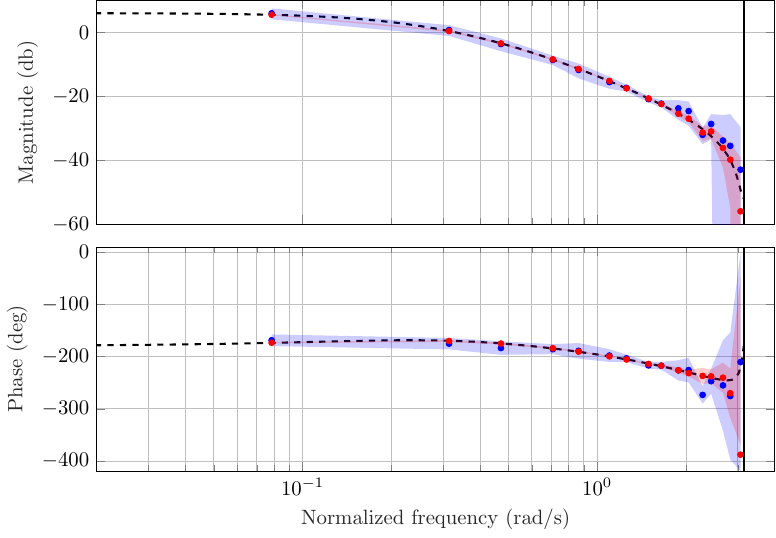}
    \caption{Estimated FRF of the true system (\kern0.5pt\protect\tikz[scale=0.6]{\protect\draw[dashed] (0,0.1)--(0.55,0.1);\protect\draw[color=white] (0,0)--(0.55,0)}) using $P=2$ periods ({\color{blue}$\bm{\cdot}$}) and $P=50$ periods ({\color{red}$\bm{\cdot}$}) along with their respective $99\%$ confidence intervals (\protect\tikz\protect\filldraw[fill=blue!20!,draw=none] (0,0) rectangle (0.2,0.2);/\protect\tikz\protect\filldraw[fill=red!20!,draw=none] (0,0) rectangle (0.2,0.2);).}
    \label{fig:frf}
\end{figure}

We generate offline data using a multi-sine excitation for the injected signal $\{d_k\}_{k\in\mathbb{N}}$, which contains $M=16$ frequencies, i.e., $\bm{\omega} = \{0.0785,\allowbreak 0.3142,\allowbreak 0.4712,\allowbreak 0.7069,\allowbreak 0.8639,\allowbreak 1.0996,\allowbreak 1.2566,\allowbreak 1.4923,\allowbreak 1.6493,\allowbreak 1.8850,\allowbreak 2.0420,\allowbreak 2.2777,\allowbreak 2.4347,\allowbreak      2.6703,\allowbreak 2.8274,\allowbreak 3.0631\}$, and zero-mean Gaussian noise $n_k$ with standard deviation $0.1$. After measuring $u_k$, $d_k$ and $y_k$ for $P$ periods of the multi-sine, we compute per period the DFT of $d_k$, $u_k$ and $y_k$ at the excited frequencies to obtain $\hat{D}^d_p(\omega_m)$, $\hat{U}^d_p(\omega_m)$ and $\hat{Y}^d_p(\omega_m)$, $m\in\mathcal{M}\coloneqq \mathbb{N}_{[1,M]}$ and $p\in\mathcal{P}\coloneqq\mathbb{N}_{[1,P]}$. Using the fact that (assuming no transient phenomena and no noise),
\begin{equation}
    \hat{U}^d_p(\omega_m) = ({1-C(e^{j\omega_m})G(e^{j\omega_m})})^{-1}\hat{D}^d_p(\omega_m),\label{eq:Udp}
\end{equation}
\begin{equation}
    \hat{Y}^d_p(\omega_m) = G(e^{j\omega_m})\hat{U}^d_p(\omega_m),\label{eq:Ydp}
\end{equation}
for $m\in\mathcal{M}$ and $p\in\mathcal{P}$, we estimate, for each measured period, the transfer function at $\omega_m$ by
\begin{equation*}
    \hat{G}_p(\omega_m) = \frac{\hat{Y}^d_p(\omega_m)(\hat{D}^d_p(\omega_m))^*}{\hat{U}^d_p(\omega_m)(\hat{D}^d_p(\omega_m))^*}.
\end{equation*}
Averaging over all periods yields our estimated FRF, i.e.,
\begin{equation*}
    \hat{G}(\omega_m) = \frac{1}{P}\sum_{p\in\mathcal{P}} G_p(\omega_m).
\end{equation*}
To characterize the uncertainty on our estimated FRF, we also compute the variance of $\hat{G}(\omega_m)$ according to\footnote{More advanced methods, such as the local polynomial method, exist to estimate FRFs and their variances~\cite[see][]{Pintelon2012}. However, this simple approach suffices to illustrate our results.}
\begin{equation*}
    \operatorname{var}~\hat{G}(\omega_m) = \frac{1}{P(P-1)}\sum_{p\in\mathcal{P}} |\hat{G}_p(\omega_m)-\hat{G}(\omega_m)|^2.
\end{equation*}
The resulting FRF measurements are shown in Fig.~\ref{fig:frf} along with the associated $99\%$ confidence intervals~\cite[][Chapter 6]{vanBerkel2015} using $P=2$ and $P=50$ periods. We observe that, particularly for $P=2$, there is significant uncertainty at low frequencies due to transient phenomena, which were not accounted for in~\eqref{eq:Udp} and~\eqref{eq:Ydp} but were present in the data, and at high frequencies due to the presence of noise. Fig.~\ref{fig:frf} shows that the estimated FRF for $P=50$ is more accurate than the estimate obtained with $P=2$, due to the random noise being averaged out. This causes the uncertainty in the FRF measurements reduces when we measure more periods. In DeePC, measuring more periods (i.e., performing longer experiments) results in higher computational complexity, as discussed in Remark~\ref{rem:complexity}. However, the number of frequencies in both FRF measurements are the same and, thereby, the number of decision variables in the corresponding FreePC schemes are the same as well. This illustrates an important advantage that frequency-domain techniques, such as FreePC, have over their time-domain counterparts. 

\begin{figure}[!t]
    \centering
    \includegraphics[width=.96\linewidth]{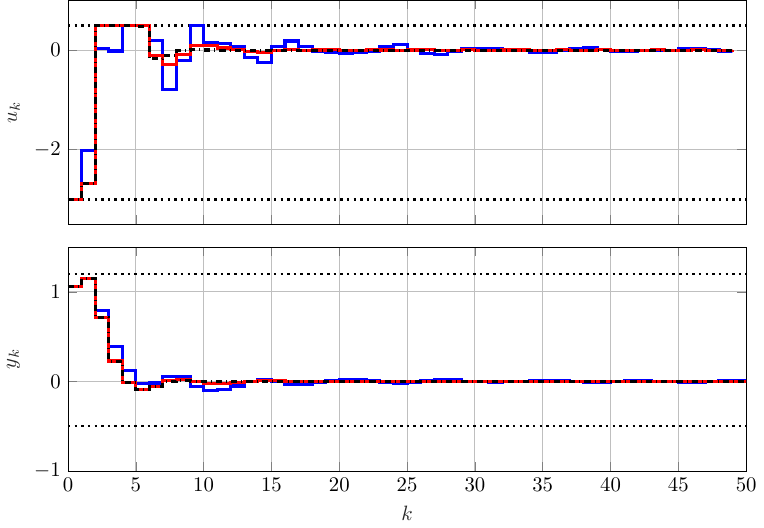}
    \caption{Comparison of FreePC using data containing $P=2$ periods (\kern0.5pt\protect\tikz[scale=0.6]{\protect\draw[color=blue, line width=1.2pt] (0,0.1)--(17pt,0.1);\protect\draw[color=white] (0,0)--(0.55,0)}) and $P=50$ periods (\kern0.5pt\protect\tikz[scale=0.6]{\protect\draw[color=red, line width=1.2pt] (0,0.1)--(0.55,0.1);\protect\draw[color=white] (0,0)--(17pt,0)}) along with a model-based benchmark (\kern0.5pt\protect\tikz[scale=0.6]{\protect\draw[color=black, dashdotted, line width=1.2pt] (0,0.1)--(17pt,0.1);\protect\draw[color=white] (0,0)--(0.55,0)}).}
    \label{fig:simulation}
\end{figure}

\begin{table}[!bt]
\centering
\caption{Monte Carlo study of FreePC using $1000$ data sets containing $P$ periods.}
\label{tab:case-study}%
\begin{tabular}{@{}c|cc@{}}
\hline
& Mean $J$ & Variance $J$\\ \hline
Model-based benchmark & $3.1801$ & -\\
$P=5$ & $3.2350$ & $3.7525 \cdot 10^{-3}$\\
$P=10$ & $3.2043$ & $3.6077\cdot 10^{-4}$\\
$P=25$ & $3.1945$ & $1.3539\cdot 10^{-4}$\\
$P=50$ &  $3.1924$ & $1.4093\cdot 10^{-4}$
\end{tabular}
\vspace*{-.1cm}
\end{table}
Next, we implement FreePC based on the frequency-domain offline data collected as described above. To this end, we incorporate our FRF measurements at the frequencies $\bm{\omega}$ into the offline data used in~\eqref{eq:FreePC} as explained in Remark~\ref{rem:frf-meas}, i.e., we take $\bm{U}^d = \{1,1,\hdots,1\}$ and $\bm{Y}^d = \{\hat{G}(\omega_1),\hat{G}(\omega_2),\hdots,\hat{G}(\omega_M)\}$. It is straightforward to verify that $\bm{U}^d$ is PE of order $2M=32$. We use a prediction horizon of $T=10$ and initial input/output data of length $\bar{T}=3n_x=6$. Moreover, we use stage cost $\ell(y,u)=y^\top Qy + u^\top Ru$ with $Q= 1$ and $R = 0.01$, regularization parameters $\lambda_g = 0.1$ and $\lambda_\sigma = 1\cdot 10^5$ and sets of admissible inputs and outputs $\mathbb{U}=[-3,0.5]$ and $\mathbb{Y} = [-0.5,1.2]$. We simulate the resulting FreePC based on the two sets of FRF measurements in Fig.~\ref{fig:frf}, obtained by measuring $P=2$ and $P=50$ periods. The resulting input and output trajectories are shown in Fig.~\ref{fig:simulation} \new{along with the trajectories obtained by using a similarly-tuned MPC scheme (based on the exact model), which serves as a benchmark.} As expected, we observe that FreePC based on $P=50$ periods performs significantly better due to the estimated FRF being more accurate, as discussed before, and, in fact, the resulting trajectories are similar to those of the model-based benchmark. To make this more insightful, we perform a Monte Carlo study, where we implement FreePC based on data sets generated as discussed before. We considered data sets containing $P\in\{5,10,25,50\}$ periods, for which we compute the cost
\begin{equation*}
    J = \textstyle\sum_{{k\in\mathbb{N}_{[0,L_{\mathrm{sim}}-1]}}} \ell(u_k,y_k),
\end{equation*}
throughout the duration $L_{\mathrm{sim}}=50$ of the simulation. The average cost and the variance of the cost throughout $1000$ runs are recorded in Table~\ref{tab:case-study} along with the cost achieved by a similarly-tuned MPC scheme (based on the exact model), which serves as a benchmark. As expected, since the FRF measurements become more representative of the true system~\eqref{eq:example-G} as the number of periods increases, we see that the average achieved cost and its variance decrease for higher $P$. Interestingly, the obtained performance appears to plateau when using sufficiently many periods and does not quite reach the achieved cost in the model-based benchmark. \new{This may be the result of the measurement noise that affects past output measurements used to estimate the internal system state, or by potential bias induced by the adopted regularization scheme, weights or the slack variables. The properties of the regularization scheme are a topic of further research.}
\section{Conclusions}\label{sec:conclusions}
In this paper, we have presented a frequency-domain data-driven predictive control scheme \new{based on a novel version of WFL that uses frequency-domain data of the to-be-controlled system}. In doing so, we bridge the gap between recent advances in data-driven predictive control and existing tools and expertise on frequency-domain control/identification that has accumulated, particularly in industry, over decades of working with classical control techniques, such as PID control and loop shaping. The FreePC scheme, which is formulated directly in terms of frequency-domain data, e.g., in the form of frequency-response-function measurements, is shown to be equivalent to the celebrated DeePC scheme. Finally, we showcased some benefits of FreePC in a numerical case study. 

\bibliography{phd-bibtex}             

\begin{thebibliography}{27}
\providecommand{\natexlab}[1]{#1}
\providecommand{\url}[1]{\texttt{#1}}
\providecommand{\urlprefix}{URL }
\expandafter\ifx\csname urlstyle\endcsname\relax
  \providecommand{\doi}[1]{doi:\discretionary{}{}{}#1}\else
  \providecommand{\doi}{doi:\discretionary{}{}{}\begingroup \urlstyle{rm}\Url}\fi

\bibitem[{{Alsati} et~al.(2023){Alsati}, {Lopez}, {Berberich}, {Allg\"{o}wer}, and {M\"{u}ller}}]{Alsati2023}
{Alsati}, M., {Lopez}, V.G., {Berberich}, J., {Allg\"{o}wer}, F., and {M\"{u}ller}, M.A. (2023).
\newblock Data-driven nonlinear predictive control for feedback linearizable systems.
\newblock \emph{IFAC-PapersOnLine}, 56(2), 617--624.

\bibitem[{{Berberich} and {Allg\"{o}wer}(2020)}]{Berberich2020}
{Berberich}, J. and {Allg\"{o}wer}, F. (2020).
\newblock A trajectory-based framework for data-driven system analysis and control.
\newblock In \emph{Eur. Control Conf.}, 1365--1370.

\bibitem[{{Berberich} et~al.(2021){Berberich}, {K\"{o}hler}, {M\"{u}ller}, and {Allg\"{o}wer}}]{Berberich2021}
{Berberich}, J., {K\"{o}hler}, J., {M\"{u}ller}, M.A., and {Allg\"{o}wer}, F. (2021).
\newblock Data-driven model predictive control with stability and robustness guarantees.
\newblock \emph{IEEE Trans. Autom. Control}, 66(4).

\bibitem[{{Berberich} et~al.(2022){Berberich}, {K\"{o}hler}, {M\"{u}ller}, and {Allg\"{o}wer}}]{Berberich2022b}
{Berberich}, J., {K\"{o}hler}, J., {M\"{u}ller}, M.A., and {Allg\"{o}wer}, F. (2022).
\newblock Linear tracking {MPC} for nonlinear systems--{Part II: The} data-driven case.
\newblock \emph{IEEE Trans. Autom. Control}, 67(9), 4406--4421.

\bibitem[{{Breschi} et~al.(2023){Breschi}, {Chiuso}, and {Formentin}}]{Breschi2023}
{Breschi}, V., {Chiuso}, A., and {Formentin}, S. (2023).
\newblock Data-driven predictive control in a stochastic setting: {A} unified framework.
\newblock \emph{Automatica}, 152, 110961.

\bibitem[{{Burgos} et~al.(2014){Burgos}, {L\'{o}pez Mart\'{i}nez}, {van de Molengraft}, and {Steinbuch}}]{Burgos2014}
{Burgos}, J.G., {L\'{o}pez Mart\'{i}nez}, C.A., {van de Molengraft}, R., and {Steinbuch}, M. (2014).
\newblock Frequency domain tuning method for unconstrained linear output feedback model predictive control.
\newblock In \emph{Proc. 19th IFAC World Congress}, 7455--7460.

\bibitem[{{Colin} et~al.(2020){Colin}, {Bombois}, {Bako}, and {Morelli}}]{Colin2020}
{Colin}, K., {Bombois}, X., {Bako}, L., and {Morelli}, F. (2020).
\newblock Data informativity for the open-loop identification of {MIMO} systems in the prediction error framework.
\newblock \emph{Automatica}, 117, 109000.

\bibitem[{{Coulson} et~al.(2019){Coulson}, {Lygeros}, and {D\"{o}rfler}}]{Coulson2019}
{Coulson}, J., {Lygeros}, J., and {D\"{o}rfler}, F. (2019).
\newblock Data-enabled predictive control: {In} the shallows of the {DeePC}.
\newblock In \emph{Eur. Control Conf.}, 307--312.

\bibitem[{{Faulwasser} et~al.(2023){Faulwasser}, {Ou}, {Pan}, {Schmitz}, and {Worthmann}}]{Faulwasser2023}
{Faulwasser}, T., {Ou}, R., {Pan}, G., {Schmitz}, P., and {Worthmann}, K. (2023).
\newblock Behavioral theory for stochastic systems? {A} data-driven journey from {Willems} to {Wiener} and back again.
\newblock \emph{Annu. Rev. Control}, 55, 92--117.

\bibitem[{{Ferizbegovic} et~al.(2021){Ferizbegovic}, {Hjalmarsson}, {Mattsson}, and {Sch\"{o}n}}]{Ferizbegovic2021}
{Ferizbegovic}, M., {Hjalmarsson}, H., {Mattsson}, P., and {Sch\"{o}n}, T.B. (2021).
\newblock {Willems}' fundamental lemma based on second-order moments.
\newblock In \emph{60th IEEE Conf. Decis. Control}, 396--401.

\bibitem[{{Hespanha}(2018)}]{Hespanha2018}
{Hespanha}, J.P. (2018).
\newblock \emph{Linear systems theory}.
\newblock Princeton University Press.

\bibitem[{{Lazar}(2023)}]{Lazar2023}
{Lazar}, M. (2023).
\newblock Basis functions nonlinear data-enabled predictive control: {Consistent} and computationally efficient formulations.
\newblock Preprint: \url{https://arxiv.org/abs/2311.05360}.

\bibitem[{{Markovsky} and {Ossareh}(2024)}]{Markovsky2024}
{Markovsky}, I. and {Ossareh}, H. (2024).
\newblock Finite-data nonparametric frequency response evaluation without leakage.
\newblock \emph{Automatica}, 159, 111351.

\bibitem[{{Mayne}(2014)}]{Mayne2014}
{Mayne}, D.Q. (2014).
\newblock Model predictive control: {Recent} developments and future promise.
\newblock \emph{Automatica}, 50, 2967--2986.

\bibitem[{{Meijer} et~al.(2023){Meijer}, {Nouwens}, {Dolk}, and {Heemels}}]{Meijer2024-ecc-arxiv}
{Meijer}, T.J., {Nouwens}, S.A.N., {Dolk}, V.S., and {Heemels}, W.P.M.H. (2023).
\newblock A frequency-domain version of {Willems'} fundamental lemma.
\newblock Preprint: \url{https://arxiv.org/abs/2311.15284}.

\bibitem[{{\"{O}zkan} et~al.(2012){\"{O}zkan}, {Meijs}, and {Backx}}]{Ozkan2012}
{\"{O}zkan}, L., {Meijs}, J., and {Backx}, A.C.P.M. (2012).
\newblock A frequency domain approach for {MPC} tuning.
\newblock \emph{Comput. Aided Chem. Eng.}, 31, 1632--1636.

\bibitem[{{Pan} et~al.(2023){Pan}, {Ou}, and {Faulwasser}}]{Pan2023}
{Pan}, G., {Ou}, R., and {Faulwasser}, T. (2023).
\newblock Towards data-driven stochastic predictive control.
\newblock \emph{Int. J. Robust Nonlinear Control}, 1--23.
\newblock \doi{10.1002/rnc.6812}.

\bibitem[{{Pintelon} and {Schoukens}(2012)}]{Pintelon2012}
{Pintelon}, R. and {Schoukens}, J. (2012).
\newblock \emph{System Identification: {A} Frequency Domain Approach}.
\newblock Wiley.

\bibitem[{{Sathyanarayanan} et~al.(2023){Sathyanarayanan}, {Pan}, and {Faulwasser}}]{Sathyanarayanan2023}
{Sathyanarayanan}, K.K., {Pan}, G., and {Faulwasser}, T. (2023).
\newblock Towards data-driven predictive control using wavelets.
\newblock \emph{IFAC-PapersOnLine}, 56(2), 632--637.

\bibitem[{{Schmitz} et~al.(2022){Schmitz}, {Faulwasser}, and {Worthmann}}]{Schmitz2022}
{Schmitz}, P., {Faulwasser}, T., and {Worthmann}, K. (2022).
\newblock Willems' fundamental lemma for linear descriptor systems and its use for data-driven output-feedback {MPC}.
\newblock \emph{IEEE Control Syst. Lett.}, 6, 2443--2448.

\bibitem[{{Shah} and {Engell}(2013)}]{Shah2013}
{Shah}, G. and {Engell}, S. (2013).
\newblock Multivariable {MPC} design based on a frequency response approximation approach.
\newblock In \emph{Eur. Control Conf}, 13--18.

\bibitem[{{S\"{o}derstr\"{o}m} and {Stoica}(1989)}]{Soderstrom1989}
{S\"{o}derstr\"{o}m}, T. and {Stoica}, P. (1989).
\newblock \emph{System identification}.
\newblock Prentice Hall.

\bibitem[{{van Berkel}(2015)}]{vanBerkel2015}
{van Berkel}, M. (2015).
\newblock \emph{Estimation of heat transport coefficients in fusion plasmas}.
\newblock Ph.D. thesis, Technische Universiteit Eindhoven.

\bibitem[{{van Waarde} et~al.(2020){van Waarde}, {De Persis}, {Camlibel}, and {Tesi}}]{vanWaarde2020}
{van Waarde}, H.J., {De Persis}, C., {Camlibel}, M.K., and {Tesi}, P. (2020).
\newblock Willems' fundamental lemma for state-space systems and its extension to multiple datasets.
\newblock \emph{IEEE Control Syst. Lett.}, 4(3), 602--607.

\bibitem[{{Verheijen} et~al.(2023){Verheijen}, {Breschi}, and {Lazar}}]{Verheijen2023}
{Verheijen}, P.C.N., {Breschi}, V., and {Lazar}, M. (2023).
\newblock Handbook of linear data-driven predictive control: {Theory}, implementation and design.
\newblock \emph{Annu. Rev. Control}, 56, 100914.

\bibitem[{{Verhoek} et~al.(2021){Verhoek}, {Abbas}, {T\'{o}th}, and {Haesaert}}]{Verhoek2021b}
{Verhoek}, C., {Abbas}, H.S., {T\'{o}th}, R., and {Haesaert}, S. (2021).
\newblock Data-driven predictive control for linear parameter-varying systems.
\newblock \emph{IFAC-PapersOnLine}, 54(8), 101--108.

\bibitem[{{Willems} et~al.(2005){Willems}, {Rapisarda}, {Markovsky}, and {De Moor}}]{Willems2005}
{Willems}, J.C., {Rapisarda}, P., {Markovsky}, I., and {De Moor}, B.L.M. (2005).
\newblock A note on persistency of excitation.
\newblock \emph{Syst. Control Lett.}, 54(4), 325--329.

\end{thebibliography}

\end{document}